\documentclass{appolb}
\usepackage{epsfig}

\begin{document}
\title{$\eta$-$\eta^\prime$ mixing from $V\to P\gamma$ and $J/\psi\to VP$ decays%
\thanks{Presented at Symposium on Meson Physics (extended COSY-11 Collaboration Meeting), Cracow, Poland, 1--4 October 2008.}%
}
\author{Rafel Escribano
\address{Grup de F\'{\i}sica Te\`orica and IFAE, Universitat Aut\`onoma de Barcelona,\\
E-08193 Bellaterra (Barcelona), Spain}
}
\maketitle
\begin{abstract}
The $\eta$-$\eta^\prime$ pseudoscalar mixing angle and the gluonium content of the
$\eta^\prime$ meson are deduced from an updated phenomenological analysis of
$V\to P\gamma$ and $J/\psi\to VP$ decays.
In absence of gluonium,
the value of the mixing angle in the quark-flavour basis is found to be $\phi_P=(41.5\pm 1.2)^\circ$
from $V\to P\gamma$ and $\phi_P=(40.5\pm 2.4)^\circ$ from $J/\psi\to VP$.
In presence of gluonium,
the values for the mixing angle and the gluonic content of the $\eta^\prime$ wave function are
$\phi_P=(41.4\pm 1.3)^\circ$ and $Z^2_{\eta^\prime}=0.04\pm 0.09$ from $V\to P\gamma$ and
$\phi_P=(44.5\pm 4.3)^\circ$ and $Z^2_{\eta^\prime}=0.28\pm 0.21$ from $J/\psi\to VP$, respectively.
\end{abstract}
  
\section{Introduction}
\label{intro}
Is $\eta^\prime$ partially made of gluonium?
To answer this question we perform a phenomenological analysis of radiative
$V\to P\gamma$ and $J/\psi\to VP$ decays,
with $V=\rho, K^\ast, \omega, \phi$ and $P=\pi, K, \eta, \eta^\prime$,
in order to determine the gluonic content of the $\eta^\prime$ wave function.
Similar analyses were driven in the seminal work by Rosner \cite{Rosner:1982ey},
where the allowed gluonic admixture in the $\eta^\prime$ could not be established due to the lack of data on $\phi\to\eta^\prime\gamma$, and, later on, by Kou who pointed out that the $\eta^\prime$ gluonic component might be as large as 26\% \cite{Kou:1999tt}.
More recently, the study by C.~E.~Thomas over a large number of different processes
concludes that while the data hint at a small gluonic component in the $\eta^\prime$,
the results depend sensitively on unknown form factors associated with exclusive dynamics
\cite{Thomas:2007uy}. 

From the experimental side,
the KLOE Collaboration, combining the new measurement of 
$R_\phi\equiv B(\phi\to\eta^\prime\gamma)/B(\phi\to\eta\gamma)$ with other constraints,
has estimated the gluonium content of the $\eta^\prime$ meson as
$Z_{\eta^\prime}^2=0.14\pm 0.04$ \cite{Ambrosino:2006gk}.
This new result contrasts with the former value $Z_{\eta^\prime}^2=0.06^{+0.09}_{-0.06}$,
which was compatible with zero and consistent with a gluonium fraction below 15\%
\cite{Aloisio:2002vm}.

A more extensive version of the present work including a detailed analysis also for the case of the
$\eta$,
a comparison with other approaches,
and a complete list of references
can be found in Refs.~\cite{Escribano:2007cd,Escribano:2008rq}.

\section{Notation}
\label{notation}
We work in a basis consisting of the states \cite{Rosner:1982ey}
\begin{equation}
\label{quarkstates}
|\eta_q\rangle\equiv\frac{1}{\sqrt{2}}|u\bar u+d\bar d\rangle\ ,\qquad
|\eta_s\rangle\equiv |s\bar s\rangle\ ,\qquad
\label{gluoniumstate}
|G\rangle=|\mbox{gluonium}\rangle\ .
\end{equation}
The physical states $\eta$ and $\eta^\prime$ are assumed to be the linear combinations
\begin{equation}
\label{physicaleta}
|\eta\rangle=X_\eta |\eta_q\rangle+Y_\eta |\eta_s\rangle+Z_\eta |G\rangle\ ,\qquad
\label{physicaletap}
|\eta^\prime\rangle=
X_{\eta^\prime}|\eta_q\rangle+Y_{\eta^\prime}|\eta_s\rangle+Z_{\eta^\prime}|G\rangle\ ,
\end{equation}
with $X_{\eta (\eta^\prime)}^2+Y_{\eta (\eta^\prime)}^2+Z_{\eta (\eta^\prime)}^2=1$.
A significant gluonic admixture in a state is possible only if
$Z_{\eta (\eta^\prime)}^2=1-X_{\eta (\eta^\prime)}^2-Y_{\eta (\eta^\prime)}^2>0$.
The implicit assumptions in Eq.~(\ref{physicaleta}) are the following:
i) no mixing with $\pi^0$ ---isospin symmetry, and
ii) no mixing with radial excitations or $\eta_c$ states.
Assuming the absence of gluonium for the $\eta$,
the coefficients $X_{\eta (\eta^\prime)}$, $Y_{\eta (\eta^\prime)}$ and $Z_{\eta (\eta^\prime)}$
are described in terms of two angles (see Ref.~\cite{Escribano:2007cd} for details),
\begin{equation}
\begin{array}{c}
\label{Xetaetap}
X_\eta=\cos\phi_P\ ,\qquad X_{\eta^\prime}=\sin\phi_P\cos\phi_{\eta^\prime G}\ ,\\[1ex]
\label{Yetaetap}
Y_\eta=-\sin\phi_P\ ,\hspace{1.5em} Y_{\eta^\prime}=\cos\phi_P\cos\phi_{\eta^\prime G}\ ,\\[1ex]
\label{Zetaetap}
Z_\eta=0\ ,\hspace{4.5em}  Z_{\eta^\prime}=-\sin\phi_{\eta^\prime G}\ ,
\end{array}
\end{equation}
where $\phi_P$ is the $\eta$-$\eta^\prime$ mixing angle and
$\phi_{\eta^\prime G}$ weights the amount of gluonium in the $\eta^\prime$ wave-function.
For a comprehensive treatment of $\eta$-$\eta^\prime$ mixing in absence of gluonium
see Ref.~\cite{Feldmann:1999uf}.

\section{Phenomenological model}
\subsection{$VP\gamma$ $M1$ transitions}
\label{model}
Our model for the $VP\gamma$ $M1$ transitions is based on three characteristic ingredients: 
{\it i)}
A $VP\gamma$ magnetic dipole transition proceeds via quark or antiquark spin-flip 
amplitudes proportional to $\mu_q=e_q/2m_q$. 
This effective magnetic moment breaks $SU(3)$ in a well defined way and distinguishes photon emission from strange or non-strange quarks via $m_s > \bar{m}$;
{\it ii)}
The spin-flip $V\leftrightarrow P$ conversion amplitude has then to be corrected by the 
relative overlap between the $P$ and $V$ wave functions \cite{Bramon:2000fr};
{\it iii)}
Indeed, the OZI-rule reduces considerably the possible transitions and their respective 
$VP$ wave-function overlaps: $C_s$, $C_q$ and $C_\pi$ characterize the 
$\langle\eta_s|\phi_s\rangle$, 
$\langle\eta_q|\omega_q\rangle =\langle\eta_q|\rho\rangle$ and 
$\langle\pi|\omega_q\rangle =\langle\pi|\rho\rangle$ spatial overlaps, respectively.
Notice that distinction is made between the $|\pi\rangle$ and $|\eta_q\rangle$ 
spatial extension due to the gluon or $U(1)_A$ anomaly.

The relevant $VP\gamma$ couplings are written in terms of a
$g\equiv g_{\omega_q\pi\gamma}$ as
\begin{equation}
\label{etacouplings}
\begin{array}{c}
g_{\rho\eta^{(\prime)}\gamma}=g\,z_{q}\,X_\eta^{(\prime)}\ ,\\[2ex]
g_{\omega\eta^{(\prime)}\gamma}=
\frac{1}{3}g\left(z_q\,X_\eta^{(\prime)}\cos\phi_V+
2\frac{\bar{m}}{m_s}z_s\,Y_\eta^{(\prime)}\sin\phi_V\right)\ ,\\[2ex]
g_{\phi\eta^{(\prime)}\gamma}=
\frac{1}{3}g\left(z_q\,X_\eta^{(\prime)}\sin\phi_V-
2\frac{\bar{m}}{m_s}z_s\,Y_\eta^{(\prime)}\cos\phi_V\right)\ ,\\[2ex]
\end{array}
\end{equation}
where we have redefined $z_q\equiv C_q/C_\pi$ and $z_s\equiv C_s/C_\pi$.

\subsection{$J/\psi\to VP$ transitions}
\begin{table*}
\begin{tabular}{cc}
\hline\\[-2ex]
Process & Amplitude \\[0.5ex]
\hline\\[-2ex]
$\rho\pi$ 						& $g+e$ \\[0.5ex]
$K^{\ast +}K^-+\mbox{c.c.}$		& $g(1-s)+e(2-x)$ \\[0.5ex]
$K^{\ast 0}\bar K^0+\mbox{c.c.}$ 	& $g(1-s)-e(1+x)$ \\[0.5ex]
$\omega_q\eta$ 				& $(g+e)X_\eta+\sqrt{2}\,r g[\sqrt{2}X_\eta+Y_\eta]
                                                                                             +\sqrt{2}\,r^\prime g Z_\eta$ \\[0.5ex]
$\omega_q\eta^\prime$ 			& $(g+e)X_{\eta^\prime}
                                                                 +\sqrt{2}\,r g[\sqrt{2}X_{\eta^\prime}+Y_{\eta^\prime}]
                                                                 +\sqrt{2}\,r^\prime g Z_{\eta^\prime}$ \\[0.5ex]
$\phi_s\eta$ 					& $[g(1-2s)-2e x]Y_\eta
							+r g[\sqrt{2}X_\eta+Y_\eta]
                                                                 +r^\prime g Z_\eta$ \\[0.5ex]
$\phi_s\eta^\prime$ 				& $[g(1-2s)-2e x]Y_{\eta^\prime}
                                                                 +r g[\sqrt{2}X_{\eta^\prime}+Y_{\eta^\prime}]
                                                                 +r^\prime g Z_{\eta^\prime}$ \\[0.5ex]
$\rho\eta$ 					& $3e X_\eta$ \\[0.5ex]
$\rho\eta^\prime$ 				& $3e X_{\eta^\prime}$ \\[0.5ex]
$\omega_q\pi^0$ 				& $3e$ \\[0.5ex]
$\phi_s\pi^0$ 					& $0$ \\[0.5ex]
\hline\\[-2ex]
\end{tabular}
\caption{
General parametrization of amplitudes for $J/\psi\to VP$ decays.}
\label{tableA}
\end{table*}
The amplitudes for the $J/\psi\to VP$ decays are expressed in terms of an
$SU(3)$-symmetric coupling strength $g$ (SOZI amplitude)
which comes from a three-gluon annihilation diagram,
an electromagnetic coupling strength $e$ (with phase $\theta_e$ relative to $g$)
which comes from the electromagnetic interaction diagram \cite{Kowalski:1976mc},
an $SU(3)$-symmetric coupling strength which is written by $g$ with suppression factor $r$ contributed from the doubly disconnected diagram (nonet-symmetry-breaking DOZI amplitude) \cite{Haber:1985cv},
where the vector and the pseudoscalar exchange an extra gluon,
and $r^\prime$ which is the relative gluonic production amplitude representing the diagram connected to a pure glueball state.
The $SU(3)$ violation is accounted for by a factor $(1-s)$ for every strange quark contributing to $g$
and a factor $x$ for a strange quark contributing to $e$.
The general parametrization of amplitudes for $J/\psi\to VP$ decays is written in Table \ref{tableA}.

\section{Results}
\subsection{$V\to P\gamma$ analysis}
We proceed to fit our theoretical expressions for the amplitudes 
comparing the available experimental information on 
$\Gamma (V\rightarrow P\gamma)$ and $\Gamma (P\rightarrow V\gamma)$
taken exclusively from Ref.~\cite{Yao:2006px}.
In the following, we leave the $z$'s free and allow for gluonium in the $\eta^\prime$ wave function only.
This will permit us to fix the gluonic content of the $\eta^\prime$ in a way identical to the experimental measurement by KLOE.
However, as a matter of comparison, we first consider the absence of gluonium in both mesons,
\emph{i.e.~}$\phi_{\eta G}=\phi_{\eta^\prime G}=0$.
The result of the fit gives $\chi^2/$d.o.f.=4.4/5 with
\begin{equation}
\label{zphiP}
\begin{array}{c}
g=0.72\pm 0.01\ \mbox{GeV$^{-1}$}\ ,\quad \phi_P=(41.5\pm 1.2)^\circ\ ,\quad
\phi_V=(3.2\pm 0.1)^\circ\ ,\\[2ex]
\frac{m_s}{\bar m}=1.24\pm 0.07\ ,\quad
z_q=0.86\pm 0.03\ ,\quad z_s=0.78\pm 0.05\ .
\end{array}
\end{equation}

Now we assume $\phi_{\eta G}=0$, \textit{i.e.}~$Z_\eta=0$
and proceed to fit the gluonic content of the $\eta^\prime$ wave function under this assumption.
The results of the new fit are\footnote{
There is a sign ambiguity in $\phi_{\eta^\prime G}$ that cannot be decided since this angle
enters into $X_{\eta^\prime}$ and $Y_{\eta^\prime}$ through a cosine.}
\begin{equation}
\label{zphiPphietapG}
\begin{array}{c}
g=0.72\pm 0.01\ \mbox{GeV$^{-1}$}\ ,\quad \frac{m_s}{\bar m}=1.24\pm 0.07\ ,\quad
\phi_V=(3.2\pm 0.1)^\circ\ ,\\[2ex]
\phi_P=(41.4\pm 1.3)^\circ\ ,\quad |\phi_{\eta^\prime G}|=(12\pm 13)^\circ\ ,\\[2ex]
z_q=0.86\pm 0.03\ ,\quad z_s=0.79\pm 0.05\ ,
\end{array}
\end{equation}
with $\chi^2/$d.o.f.=4.2/4.
The result obtained for $\phi_{\eta^\prime G}$ suggests a very small amount of gluonium in the
$\eta^\prime$ wave function, 
$|\phi_{\eta^\prime G}|=(12\pm 13)^\circ$ or $Z_{\eta^\prime}^2=0.04\pm 0.09$.
Our values contrast with those reported by KLOE recently, $\phi_P=(39.7\pm 0.7)^\circ$ and
$|\phi_{\eta^\prime G}|=(22\pm 3)^\circ$ ---or $Z_{\eta^\prime}^2=0.14\pm 0.04$---
\cite{Ambrosino:2006gk}.

%
\begin{figure}[t]
\centerline{\includegraphics[width=0.70\textwidth]{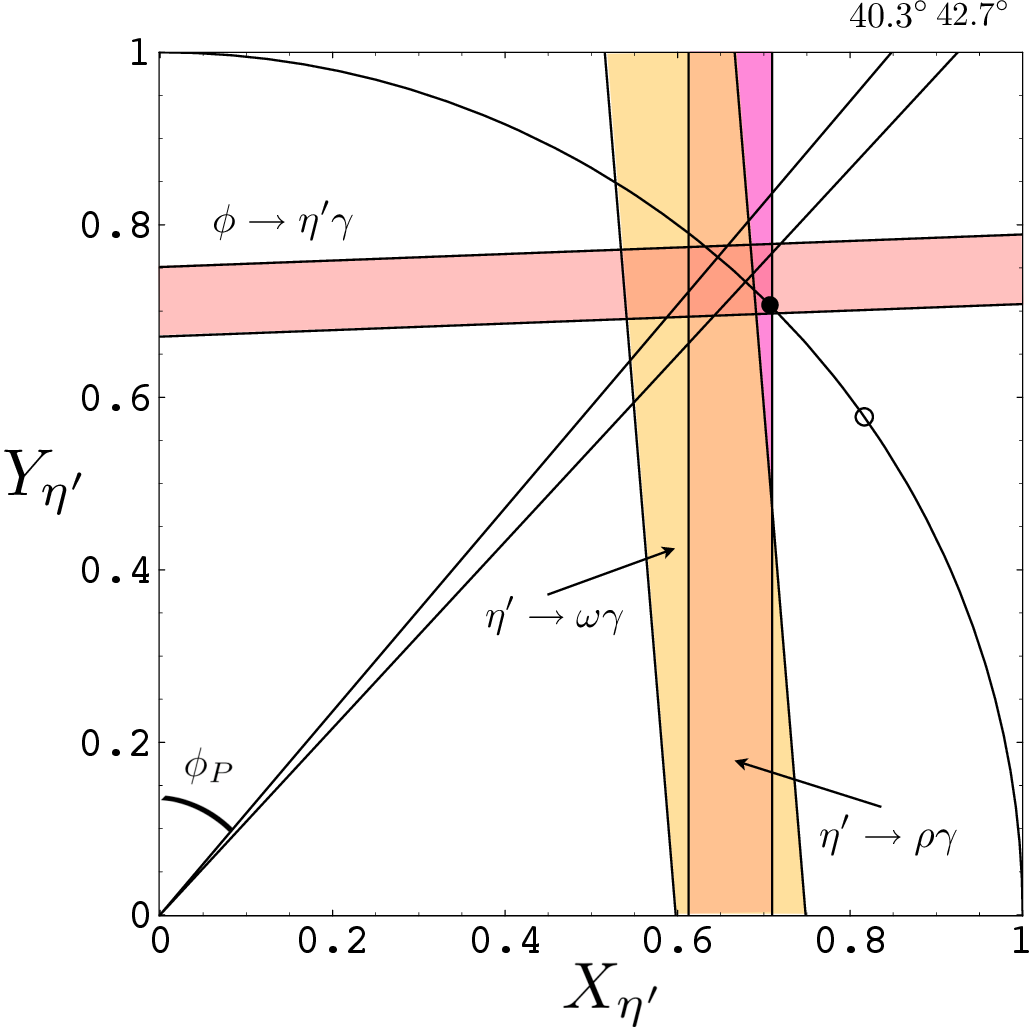}}  
\caption{Constraints on non-strange ($X_{\eta^\prime}$) and strange ($Y_{\eta^\prime}$) quarkonium mixing coefficients in the $\eta^\prime$.
The vertical and inclined bands are the regions for $X_{\eta^\prime}$ and $Y_{\eta^\prime}$ allowed by the experimental couplings of the $\eta^\prime\to(\rho,\omega)\gamma$ and
$\phi\to\eta^\prime\gamma$ transitions.}
\label{plotetap}
\end{figure}
%
Our main results can also be displayed graphically following
Refs.~\cite{Rosner:1982ey,Kou:1999tt,Ambrosino:2006gk}.
In Fig.~\ref{plotetap}, we plot the regions for the $X_{\eta^\prime}$ and $Y_{\eta^\prime}$ parameters which are allowed by the experimental couplings of the
$\eta^\prime\to\rho\gamma$, $\eta^\prime\to\omega\gamma$ and $\phi\to\eta^\prime\gamma$ transitions.
The limits of the bands are given at 68\% CL or $1\sigma$.
The remaining parameters are taken from Eq.~(\ref{zphiPphietapG}).
In addition to the bands, we have also plotted the circular boundary denoting the constraint 
$X_{\eta^\prime}^2+Y_{\eta^\prime}^2\leq 1$ as well as the favoured region for the
$\eta$-$\eta^\prime$ mixing angle assuming the \emph{absence of gluonium},
$40.3^\circ\leq\phi_P\leq 42.7^\circ$, obtained at $1\sigma$
from the corresponding fitted value in Eq.~(\ref{zphiP}).
There exists an intersection region of the three bands inside and on the circumference.
As most of this region is interior but close to the circular boundary it may well indicate a small but non necessarily zero gluonic content of the $\eta^\prime$.
Indeed, we have found $Z_{\eta^\prime}^2=0.04\pm 0.09$ (or $|Z_{\eta^\prime}|=0.2\pm 0.2$) or using the angular description $|\phi_{\eta^\prime G}|=(12\pm 13)^\circ$.

\subsection{$J/\psi\to VP$ analysis}
Given the large number of parameters to be fitted, 13 in the most general case for 11 observables
(indeed 10 because there is only an upper limit for $\phi\pi^0$),
we perform the following simplifications.
First, we fix the parameters $x=m_{u,d}/m_s$ and the vector mixing angle $\phi_V$ to the values obtained from a recent fit to the most precise data on $V\to P\gamma$ decays \cite{Escribano:2007cd},
that is $m_s/m_{u,d}=1.24\pm 0.07$ which implies $x=0.81\pm 0.05$ and $\phi_V=(3.2\pm 0.1)^\circ$.
Second, we do not allow for gluonium in the $\eta$ wave function,
thus the mixing pattern of $\eta$ and $\eta^\prime$ is given by the mixing angle $\phi_P$ and the coefficient $Z_{\eta^\prime}$ (see Notation).

We proceed to present the results of the fits.
To describe data without considering the contribution from the doubly disconnected diagram (terms proportional to $rg$) has been shown to be unfeasible \cite{Coffman:1988ve}.
Therefore, it is required to take into account nonet-symmetry-breaking effects.
We have also tested that it is not possible to get a reasonable fit setting the $SU(3)$-breaking correction $s$ to its symmetric value, \textit{i.e.~}$s=0$.
If gluonium is not allowed in the $\eta^\prime$ wave function,
the result of the fit gives $\phi_P=(40.4\pm 2.4)^\circ$ ---or $\theta_P=(-14.3\pm 2.4)^\circ$---
with $\chi^2/\mbox{d.o.f.}=3.6/4$, in disagreement at the $2\sigma$ level with
$\theta_P=(-19.1\pm 1.4)^\circ$ \cite{Jousset:1988ni},
$\theta_P=(-19.2\pm 1.4)^\circ$ \cite{Coffman:1988ve}, and
$\theta_P\simeq -20^\circ$ \cite{Morisita:1990cg},
but in correspondence with $\phi_P=(39.9\pm 2.9)^\circ$ \cite{Feldmann:1998vh}
and $\phi_P=(40\pm 2)^\circ$ \cite{Thomas:2007uy}.
In some analyses, $x$ is kept fixed to one since it always appears multiplying $e$ and hence also considered as a second order contribution.
In this case, our fit gives $\phi_P=(40.2\pm 2.4)^\circ$ with $\chi^2/\mbox{d.o.f.}=3.4/4$.
However, none of the former analyses include the effects of a vector mixing angle different from zero.
It was already noticed in Ref.~\cite{Bramon:1997mf} that these effects,
which were considered there for the first time, turn out to be crucial to find a less negative value of the
$\eta$-$\eta^\prime$ mixing angle.
If we take now the fitted value $\phi_V=+3.2^\circ$ (see above), one gets
$\phi_P=(40.5\pm 2.4)^\circ$ with $\chi^2/\mbox{d.o.f.}=4.2/4$ and
$\phi_P=(40.3\pm 2.4)^\circ$ with $\chi^2/\mbox{d.o.f.}=3.8/4$ for $x=0.81$ and $x=1$, respectively.
These new fits seem to refute the strong correlation between the two mixing angles found in
Ref.~\cite{Bramon:1997mf}.
One interesting feature of the present analysis is the effect produced in the fits by the new averaged value of the $\rho\pi$ branching ratio.
For instance, if $B(\rho\pi)=(16.9\pm 1.5)\%$ \cite{Amsler:2008zz} is replaced by its old value
$(12.8\pm 1.0)\%$ \cite{Barnett:1996hr} one gets $\phi_P=(37.7\pm 1.5)^\circ$
with $\chi^2/\mbox{d.o.f.}=8.8/4$ for $x=0.81$ and $\phi_V=3.2^\circ$, 
\textit{i.e.}~the central value and the error of the mixing angle become smaller and the quality of the fit worse.
However, this value is now in agreement with that found in Ref.~\cite{Bramon:1997mf}.

As stated, the former fits are performed assuming the absence of gluonium in $\eta^\prime$.
Now, we redo some of the fits accepting a gluonic content in the $\eta^\prime$ wave function.
For $\phi_V=0$, the values
$\phi_P=(44.8\pm 4.3)^\circ$ and $Z^2_{\eta^\prime}=0.29\pm 0.21$ with $\chi^2/\mbox{d.o.f.}=2.3/2$ and
$\phi_P=(45.0\pm 4.3)^\circ$ and $Z^2_{\eta^\prime}=0.30\pm 0.20$ with $\chi^2/\mbox{d.o.f.}=1.9/2$
are obtained for $x=0.81$ and $x=1$, respectively.
For $\phi_V=+3.2^\circ$, one gets
$\phi_P=(44.5\pm 4.3)^\circ$ and $Z^2_{\eta^\prime}=0.28\pm 0.21$ with $\chi^2/\mbox{d.o.f.}=3.0/2$ and
$\phi_P=(44.6\pm 4.3)^\circ$ and $Z^2_{\eta^\prime}=0.30\pm 0.21$ with $\chi^2/\mbox{d.o.f.}=2.6/2$, respectively.
These fits seem to favour a substantial gluonic component in $\eta^\prime$ which is, however, compatible with zero at $2\sigma$ due to the large uncertainty.
In all cases, the mixing angle and most of the other parameters are consistent with those assuming no gluonium but with larger uncertainties due to fewer constraints.
The parameter $r^\prime$ weighting the relative gluonic production amplitude is consistent with zero and has a large uncertainty.
These results are in agreement with
the values $\phi_P=(45\pm 4)^\circ$ and $Z^2_{\eta^\prime}=0.30\pm 0.21$ 
---or $\phi_{\eta^\prime G}=(33\pm 13)^\circ$---
found in Ref.~\cite{Thomas:2007uy}.

\section{Conclusions}
In this work
we have performed an updated phenomenological analysis of an accurate and exhaustive set of 
$V\to P\gamma$ and $J/\psi\to VP$ decays with the purpose of determining the quark and gluon content of the $\eta$ and $\eta^\prime$ mesons.

Our conclusions from the $V\to P\gamma$ analysis are the following.
First, accepting the absence of gluonium for the $\eta$ meson, the current experimental data on
$VP\gamma$ transitions indicate within our model a negligible gluonic content for the
$\eta^\prime$ meson, $Z_{\eta^\prime}^2=0.04\pm 0.09$.
Second, this gluonic content of the 
$\eta^\prime$ wave function amounts to $|\phi_{\eta^\prime G}|=(12\pm 13)^\circ$ and the
$\eta$-$\eta^\prime$ mixing angle is found to be $\phi_P=(41.4\pm 1.3)^\circ$.
Third, imposing the absence of gluonium for both mesons one finds $\phi_P=(41.5\pm 1.2)^\circ$,
in agreement with the former result.
Finally, we would like to stress that more refined experimental data, particularly for the
$\phi\to\eta^\prime\gamma$ channel, will contribute decisively to clarify this issue.

Our conclusions from the $J/\psi\to VP$ analysis follow.
First, assuming the absence of gluonium,
the $\eta$-$\eta^\prime$ mixing angle is found to be $\phi_P=(40.5\pm 2.4)^\circ$,
in agreement with recent experimental measurements \cite{Ambrosino:2006gk} and
phenomenological estimates \cite{Thomas:2007uy}.
Second, if gluonium is allowed in the $\eta^\prime$ wave function,
the values obtained are $\phi_P=(44.5\pm 4.3)^\circ$ and $Z^2_{\eta^\prime}=0.28\pm 0.21$
---or $|\phi_{\eta^\prime G}|=(32\pm 13)^\circ$,
which suggest within our model a substantial gluonic component in $\eta^\prime$.
Third, the inclusion of vector mixing angle effects, not included in previous analyses,
turns out to be irrelevant.
Finally, it is worth noticing that the recent reported values of $B(J/\psi\to\rho\pi)$
by the BABAR \cite{Aubert:2004kj} and BES \cite{Bai:2004jn} Collab.~are crucial to obtain a consistent description of data.

\section*{Acknowledgements}
I would like to express my gratitude to the Organizing Committee (and, in particular, to P.~Moskal)
for the opportunity of presenting this contribution, and for the pleasant
and interesting symposium we have enjoyed.
This work was supported in part by the Ramon y Cajal program,
the Ministerio de Educaci\'on y Ciencia under grant CICYT-FEDER-FPA2008-01430,
the EU Contract No.~MRTN-CT-2006-035482, ``FLAVIAnet'',
the Spanish Consolider-Ingenio 2010 Programme CPAN (CSD2007-00042), and
the Generalitat de Catalunya under grant 2005-SGR-00994.


\begin{thebibliography}{99}

\bibitem{Rosner:1982ey}
  J.~L.~Rosner,
  Phys.\ Rev.\  D {\bf 27} (1983) 1101.

\bibitem{Kou:1999tt}
  E.~Kou,
  Phys.\ Rev.\  D {\bf 63} (2001) 054027
  [arXiv:hep-ph/9908214].

\bibitem{Thomas:2007uy}
  C.~E.~Thomas,
  JHEP {\bf 0710} (2007) 026
  [arXiv:0705.1500 [hep-ph]].

\bibitem{Ambrosino:2006gk}
  F.~Ambrosino {\it et al.}  [KLOE Collaboration],
  Phys.\ Lett.\  B {\bf 648} (2007) 267
  [arXiv:hep-ex/0612029].
  
\bibitem{Aloisio:2002vm}
  A.~Aloisio {\it et al.}  [KLOE Collaboration],
  Phys.\ Lett.\  B {\bf 541} (2002) 45
  [arXiv:hep-ex/0206010].

\bibitem{Escribano:2007cd}
  R.~Escribano and J.~Nadal,
  JHEP {\bf 0705} (2007) 006
  [arXiv:hep-ph/0703187].

\bibitem{Escribano:2008rq}
  R.~Escribano,
  arXiv:0807.4201 [hep-ph].

\bibitem{Feldmann:1999uf}
  T.~Feldmann,
  Int.\ J.\ Mod.\ Phys.\  A {\bf 15} (2000) 159
  [arXiv:hep-ph/9907491].
  
\bibitem{Bramon:2000fr}
  A.~Bramon, R.~Escribano and M.~D.~Scadron,
  Phys.\ Lett.\  B {\bf 503} (2001) 271
  [arXiv:hep-ph/0012049].

\bibitem{Kowalski:1976mc}
  H.~Kowalski and T.~F.~Walsh,
  Phys.\ Rev.\  D {\bf 14} (1976) 852.

\bibitem{Haber:1985cv}
  H.~E.~Haber and J.~Perrier,
  Phys.\ Rev.\  D {\bf 32} (1985) 2961.

\bibitem{Yao:2006px}
  W.~M.~Yao {\it et al.}  [Particle Data Group],
  J.\ Phys.\ G {\bf 33} (2006) 1.

\bibitem{Coffman:1988ve}
  D.~Coffman {\it et al.}  [MARK-III Collaboration],
  Phys.\ Rev.\  D {\bf 38} (1988) 2695 
  [Erratum-ibid.\  D {\bf 40} (1989) 3788].

\bibitem{Jousset:1988ni}
  J.~Jousset {\it et al.}  [DM2 Collaboration],
  Phys.\ Rev.\  D {\bf 41} (1990) 1389.

\bibitem{Morisita:1990cg}
  N.~Morisita, I.~Kitamura and T.~Teshima,
  Phys.\ Rev.\  D {\bf 44} (1991) 175.
  
\bibitem{Feldmann:1998vh}
  T.~Feldmann, P.~Kroll and B.~Stech,
  Phys.\ Rev.\  D {\bf 58} (1998) 114006
  [arXiv:hep-ph/9802409].

\bibitem{Bramon:1997mf}
  A.~Bramon, R.~Escribano and M.~D.~Scadron,
  Phys.\ Lett.\  B {\bf 403} (1997) 339
  [arXiv:hep-ph/9703313].

\bibitem{Amsler:2008zz}
  C.~Amsler {\it et al.}  [Particle Data Group],
  Phys.\ Lett.\  B {\bf 667} (2008) 1.
  
\bibitem{Barnett:1996hr}
  R.~M.~Barnett {\it et al.}  [Particle Data Group],
  Phys.\ Rev.\  D {\bf 54} (1996) 1.

\bibitem{Aubert:2004kj}
  B.~Aubert {\it et al.}  [BABAR Collaboration],
  Phys.\ Rev.\  D {\bf 70} (2004) 072004 
  [arXiv:hep-ex/0408078].

\bibitem{Bai:2004jn}
  J.~Z.~Bai {\it et al.}  [BES Collaboration],
  Phys.\ Rev.\  D {\bf 70} (2004) 012005
  [arXiv:hep-ex/0402013].

\end{thebibliography}
\end{document}